\documentclass[aps,pra,twocolumn,superscriptaddress,showpacs,nofootinbibfloatfix,amsmath,amsfonts,amssymb]{revtex4-2}%
\usepackage{amsmath,amsfonts,amssymb,color}
\usepackage{amsthm}
\usepackage{leftidx}
\usepackage{graphicx}
\usepackage{xcolor}
\usepackage{dcolumn}
\usepackage{bm}
\usepackage{epstopdf}
\usepackage{epsfig}
\usepackage{environ}
\usepackage{pdfcomment}

\usepackage{multirow}
\usepackage{setspace}
\usepackage{color}

\usepackage{float}
\usepackage[T1]{fontenc}
\usepackage[latin9]{inputenc}
\usepackage{setspace}
\usepackage{esint}

\usepackage{wasysym}

\usepackage{lipsum}

\providecommand{\tabularnewline}{\\}

\begin{document}


\title{Topology and edge modes surviving criticality in non-Hermitian Floquet
	systems}

\author{Longwen Zhou}
\email{zhoulw13@u.nus.edu}
\affiliation{%
	College of Physics and Optoelectronic Engineering, Ocean University of China, Qingdao, China 266100
}
\affiliation{%
	Qingdao Key Laboratory of Advanced Optoelectronics, Qingdao, China 266100
}
\affiliation{%
	Engineering Research Center of Advanced Marine Physical Instruments and Equipment of MOE, Qingdao, China 266100
}

\date{\today}

\begin{abstract}
The discovery of critical points that can host quantized nonlocal
order parameters and degenerate edge modes relocate the study of symmetry-protected
topological phases (SPTs) to gapless regions. In this letter, we reveal
gapless SPTs (gSPTs) in systems tuned out-of-equilibrium by periodic
drivings and non-Hermitian couplings. Focusing on one-dimensional
models with sublattice symmetry, we introduce winding numbers by applying
the Cauchy's argument principle to generalized Brillouin zone (GBZ), yielding
unified topological characterizations and bulk-edge correspondence
in both gapped phases and at gapless critical points. The theory is
demonstrated in a broad class of Floquet bipartite lattices, unveiling
unique topological criticality of non-Hermitian Floquet origin. Our
findings identify gSPTs in driven open systems and uncover robust
topological edge modes at phase transitions beyond equilibrium.
\end{abstract}

\pacs{}
\keywords{}
\maketitle

\emph{Introduction}.--Featured by coexistent
bulk fluctuations and symmetry-protected edge modes, gSPTs blend the
physics of quantum criticality and topology into a single context
\cite{gSPT01}. The survival of topological edge modes at critical
points further allows the storage of localized information
across phase transitions \cite{gSPT02}, whose stability could be
of great use in quantum technologies. Following theoretical
progress \cite{gSPT03,gSPT04,gSPT05,gSPT06,gSPT07,gSPT08,gSPT09,gSPT10,gSPT11,gSPT12,gSPT13,gSPT14,gSPT15,gSPT16,gSPT17,gSPT18,gSPT19,gSPT20,gSPT21,gSPT22,gSPT23,gSPT24,gSPT25,gSPT26,gSPT27,gSPT28,gSPT29,gSPT30,gSPT31,gSPT32,gSPT33,gSPT34,gSPT35,gSPT36,gSPT37,gSPT38},
observations of gSPTs were made in superconducting
processors and acoustic waveguides \cite{gSPTExp01,gSPTExp02,gSPTExp03},
attracting continued interest over the years.

Floquet drivings and non-Hermitian effects are two
means to impel a system out of equilibrium. The former
could induce symmetry-breaking and topological transitions \cite{FTP01,FTP02,FTP03},
create long-range-coupled phases with large topological numbers \cite{FTP04,FTP05,FTP06},
and generate anomalous edge modes without static counterparts \cite{FTP07,FTP08,FTP09}.
The latter could enrich the symmetry classification of topological
matter \cite{BiorthogonalQM,NHTP01,NHTP02,NHTP03}, yield exceptional points with
enhanced sensitive \cite{EP01,EP02,EP03}, and lead to non-Hermitian
skin effects (NHSEs) that challenging the Bloch-band theory
\cite{NHSE01,NHSE02,NHSE03}. Their cooperation
further results in non-Hermitian
Floquet phases with unique topology \cite{NHFTPrev}, whose
characterizations were yet mostly focused on gapped cases.

This letter establishes a framework to depict the gapless bulk topology,
$(0,\pi)$ edge modes and bulk-edge correspondence at non-Hermitian
Floquet criticality. Our approach extends a recent theory of Floquet quantum criticality
\cite{ZhouFgSPT1,ZhouFgSPT2,ZhouFgSPT3} by incorporating the GBZ scheme
\cite{GBZ01,GBZ02,GBZ03}, making
it applicable to the gaplessness of both Bloch and non-Bloch Floquet
bands \cite{NHFTP01,NHFTP02,NHFTP03,FGBZ02}. The theory is demonstrated
in a set of Floquet non-Hermitian Su-Schrieffer-Heeger (NHSSH)
chains and developed more systematically in the Supplementary Material
(SM) \cite{SM}. A unified view is offered by investigating the phase
diagram, edge states, phase transitions along topological
phase boundaries and scaling of entanglement entropy (EE). Notably,
we identify nontrivial critical points and critical
edge modes originated from non-Hermitian Floquet
drivings.

\begin{figure}
	\begin{centering}
		\includegraphics[scale=0.4]{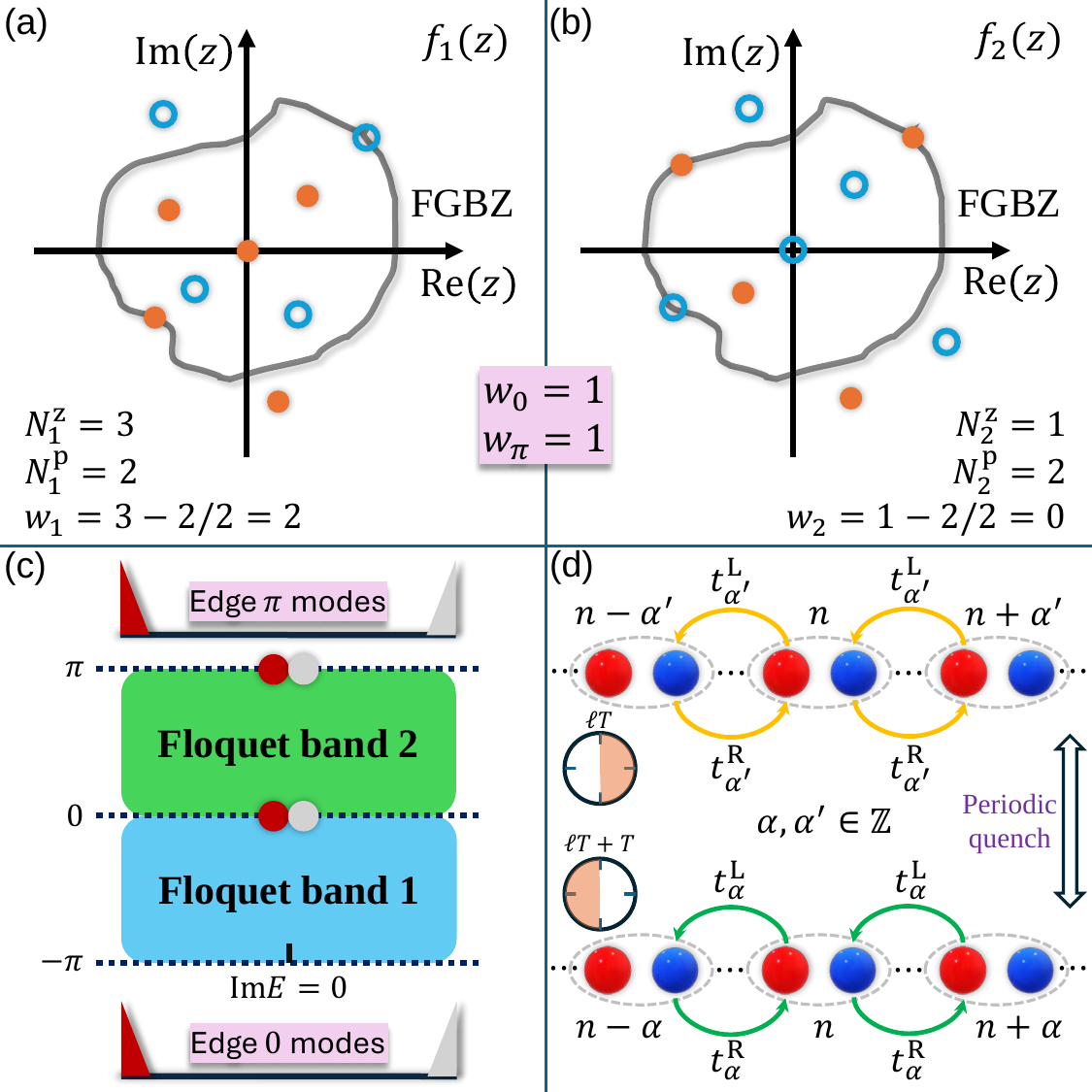}
		\par\end{centering}
	\caption{Schematic diagram. (a) and (b) show the zeros/poles
			of $f_{1,2}(z)$
			on complex plane, with the wiggled
			contour showing the FGBZ. Orange dots (cyan circles) denote the zeros
			(poles) of $f_{1,2}(z)$. The spectrum related to
			(a)--(b) is sketched in (c). The two bands are touched
			at $E=0,\pi$. The critical $0$ ($\pi$)
			edge modes are depicted by solid balls at $E=0$ ($\pi$),
			with their profiles illustrated at the bottom (top) of (c).
			(d) shows the periodically quenched NHSSH model with $\alpha,\alpha'\in{\mathbb Z}$. Each encircled set of red and blue balls denote a unit cell (indexed by $n$) with two sublattices A and B.\label{fig:Sketch}}
\end{figure}

\emph{Theory}.--We start by noting that any (non-)Hermitian
Bloch Hamiltonian of a one-dimensional (1D), two band system with sublattice symmetry
can be formally expressed as $H(k)=f(k)\sigma_{+}+g(k)\sigma_{-}$,
where $\sigma_{\pm}\equiv(\sigma_{x}\pm i\sigma_{y})/2$, $\sigma_{x,y,z}$
are Pauli matrices, and the sublattice symmetry $\Gamma=\sigma_{z}$
acts on $H(k)$ as $\Gamma H(k)\Gamma=-H(k)$. The $2\pi$-periodicity
of $H(k)$ in quasimomentum $k$ allows one to expand $f(k)$ and
$g(k)$ as polynomials of $e^{ik}$. Continuing $e^{ik}\rightarrow z$
to the whole complex plane and tracing the zero/pole locations
of polynomial $f(z)$, $g(z)$ yield a topological invariant $w$, which
could classify all gapped phases of $H(z)$ and predict a bulk-edge
correspondence consisting with the non-Bloch band theory
\cite{ZP01,GBZAL2,ZP02,ZP03,GBZAL5}. However, this approach is not directly
applicable at the gap-closing point of $H(z)$ \cite{ZhouNHgSPT},
not to mention the Floquet case. The underlying reason
is at least twofold. First, the non-Hermiticity of $H(z)$ implies
$f(z)\neq g^{*}(z)$, so that the criticality in $f(z)$
and $g(z)$ must be treated simultaneously at phase transitions instead
of being depicted by a single ratio. Second, the spectrum of a Floquet operator
could be gapless at both $0$ and $\pi$ quasienergies \cite{STF01,STF02,STF03}.
Such a ``gapless-doubling'' is of temporal origin, making it impossible
to capture its associated topology and phase transitions by a single
static Hamiltonian.

To proceed, we express the Floquet operator $U(k)=\hat{{\cal T}}e^{-\frac{i}{\hbar}\int_{t'}^{t'+T}H(k,t)dt}$
of the system in symmetric time frames \cite{STF01} by
selecting the initial time of evolution $t'$, so that the symmetrized
Floquet operator $U_{s}(k)$ unfolds the sublattice symmetry
as $\Gamma U_{s}(k)\Gamma=U_{s}^{-1}(k)$ ($s=1,2$). Here,
the Hamiltonian $H(k,t)=H(k,t+T)$ has the driving period
$T$, and $\hat{{\cal T}}$ performs time ordering. In time frame
$s$, the effective Hamiltonian of $U_{s}(k)$
reads $[U_{s}^{-1}(k)-U_{s}(k)]/(2i)$, which must have
the form $H_{s}(k)=p_{s}(k)\sigma_{+}+q_{s}(k)\sigma_{-}$ due to
its sublattice symmetry. Taking the continuation
$k\rightarrow-i\ln z$, we find a characteristic function of $H_{s}(z)$
as $f_{s}(z)=p_{s}(z)$ ($=q_{s}(z)$) if $m>m'$ ($m'>m$), where
$m$ and $m'$ are the highest positive powers of $p_{s}(z)$ and
$q_{s}(z)$ in $z$. Finding the numbers of zeros $N_{s}^{{\rm z}}$
and poles $N_{s}^{{\rm p}}$ of $f_{s}(z)$ inside the GBZ of $H_{s}(z)$ then yields
a winding number $w_{s}$ following Cauchy's argument principle, i.e.,
\begin{equation}
	w_{s}\equiv\begin{cases}
		N_{s}^{{\rm z}}-N_{s}^{{\rm p}}, & \Delta_{\pi}\neq0\\
		N_{s}^{{\rm z}}-N_{s}^{{\rm p}}/2, & \Delta_{\pi}=0
	\end{cases},\label{eq:w12main}
\end{equation}
where $\Delta_{\pi}\equiv\min_{z\in{\rm GBZ}}|E(z)-\pi|$ infers whether
$U_{s}(z)$ has a gap at quasienergy $E=\pi$ along the Floquet
GBZ (FGBZ). Combining the $(w_{1},w_{2})$ defined in two
time frames yields the topological invariants
\begin{equation}
	(w_{0},w_{\pi})=\frac{1}{2}(w_{1}+w_{2},w_{1}-w_{2})\in\mathbb{Z}\times\mathbb{Z}.\label{eq:w0Pmain}
\end{equation}
They offer a unified
characterization for all gapped and gapless non-Hermitian
Floquet phases in 1D, two-band, sublattice-symmetric systems. The
numbers of edge modes at $0$ and $\pi$ quasienergies are
related to these invariants according to the bulk-edge
correspondence
\begin{equation}
	(N_{0},N_{\pi})=2(|w_{0}|,|w_{\pi}|).\label{eq:BBCmain}
\end{equation}

We emphasize that Eqs.~(\ref{eq:w12main})--(\ref{eq:BBCmain}) are
generic, in the sense that they do not concern whether there
exists a (point or line) gap in the spectrum and whether
the underlying band theory is in Bloch or non-Bloch forms.
The only relevant factors are spatial dimensions, number of
bands and symmetries of
the system. In Hermitian limits, where FGBZ reduces to
the usual Brillouin zone, we end with the approach in
Ref.~\cite{ZhouFgSPT1}. Figs.~\ref{fig:Sketch}(a)--(c) illustrate our theory applied
to a case with gapless spectrum \cite{SM}. Below, we work out an
example to show the unique topology and edge modes at criticality
in non-Hermitian Floquet systems.

\begin{figure}
	\begin{centering}
		\includegraphics[scale=0.45]{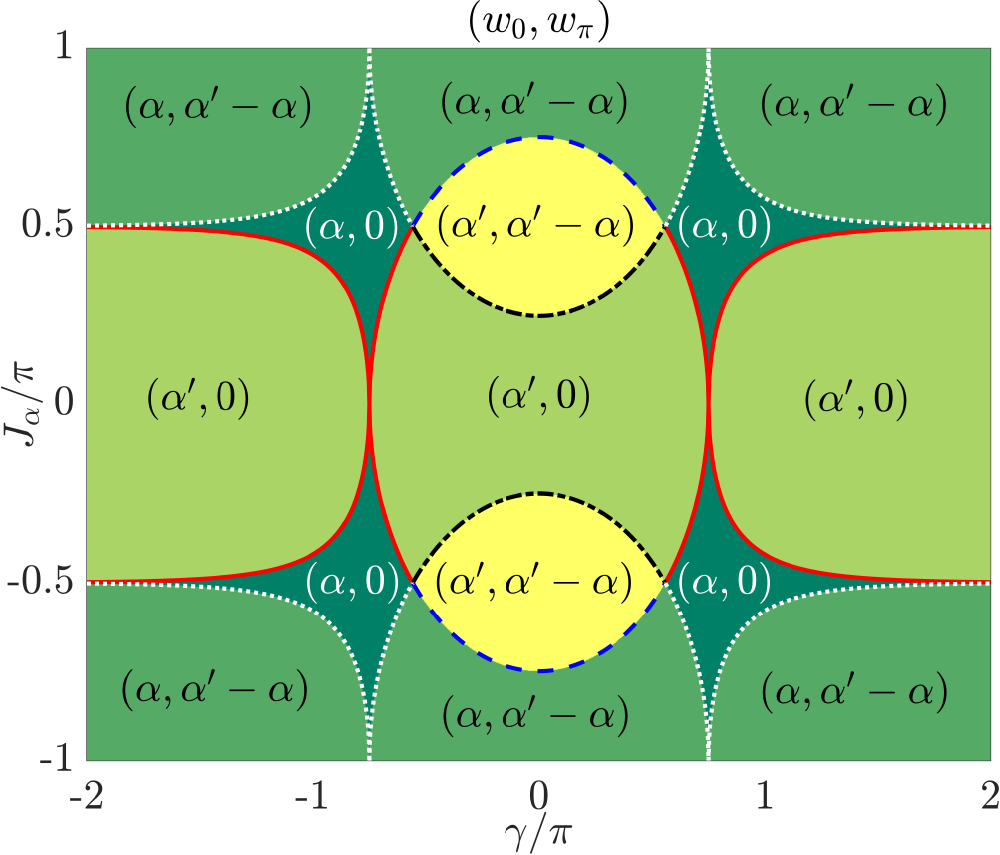}
		\par\end{centering}
	\caption{Phase diagram of the PQNHSSH $\alpha$-chain.
		Hopping amplitudes over $\alpha'$ unit cells are
		$J_{\alpha'}^{{\rm L(R)}}=J-(+)\gamma$. Regions of distinct
		colors are gapped phases, with their topological indices $(w_{0},w_{\pi})$ shown
		explicitly. The solid (dotted) lines have $(w_{0},w_{\pi})=(\alpha,0)$,
		with bulk gap closes at $E=0$ ($\pi$). The dash-dotted
		lines have $(w_{0},w_{\pi})=(\alpha',0)$, with bulk gap
		closes at $E=\pi$. The dashed lines have $(w_{0},w_{\pi})=(\alpha,\alpha'-\alpha)$,
		with bulk gap closes at $E=0$. \label{fig:PD}}
\end{figure}

\emph{Model}.--We focus on a class of periodically
quenched NHSSH (PQNHSSH) models, which may be viewed as a stacking
of SSH chains with two distinct hopping ranges in time
domain. A sketch of the model is given in
Fig.~\ref{fig:Sketch}(d), where the Floquet operator reads $\hat{U}=e^{-i\hat{{\cal H}}_{\alpha}}e^{-i\hat{{\cal H}}_{\alpha'}}$,
$\hat{{\cal H}}_{\iota}\equiv\sum_{n}(J_{\iota}^{{\rm L}}\hat{b}_{n}^{\dagger}\hat{a}_{n+\iota}+J_{\iota}^{{\rm R}}\hat{a}_{n+\iota}^{\dagger}\hat{b}_{n})$,
the hopping amplitude $J_{\iota}^{{\rm L,R}}\equiv t_{\iota}^{{\rm L,R}}T/(2\hbar)$
for $\iota=\alpha,\alpha'$, and $\hat{a}_{n}^{\dagger}$ ($\hat{b}_{n}^{\dagger}$)
creates a particle in sublattice A (B) of unit cell $n$. To evaluate
the winding numbers, we need to transform
$\hat{U}$ to symmetric time frames. Taking the initial time of evolution as
$t'=T/4$ and $3T/4$ yield Floquet operators
\begin{equation}
	\hat{{\cal U}}_{1(2)}= e^{-\frac{i}{2}\hat{{\cal H}}_{\alpha'(\alpha)}}e^{-i\hat{{\cal H}}_{\alpha(\alpha')}}e^{-\frac{i}{2}\hat{{\cal H}}_{\alpha'(\alpha)}},\label{eq:U12main}
\end{equation}
where the sublattice symmetry $\hat{\Gamma}=\sum_{n}(\hat{a}_{n}^{\dagger}\hat{a}_{n}-\hat{b}_{n}^{\dagger}\hat{b}_{n})$
is retained as $\hat{\Gamma}\hat{{\cal U}}_{s}\hat{\Gamma}=\hat{{\cal U}}_{s}^{-1}$for
$s=1,2$. This symmetry enforces the eigenstates of $\hat{U}$ to
come in pairs of opposite quasienergies $\pm E$. The Floquet
bands of $\hat{U}$ could thus meet at either $E=0$ or $\pi$,
yielding phase transitions (criticality), with the latter being of
Floquet origin. Below, we take $0\leq\alpha<\alpha'$ and
let $J_{\alpha}^{{\rm L}}=J_{\alpha}^{{\rm R}}=J_{\alpha}\in\mathbb{R}$
without losing the essence. Non-Hermitian effects then arise from nonreciprocal couplings $J_{\alpha'}^{{\rm L}}\neq(J_{\alpha'}^{{\rm R}})^{*}$.
The model we introduced here will be called the PQNHSSH $\alpha$-chain
(see Sec.~II of \cite{SM} for details).

\begin{figure}
	\begin{centering}
		\includegraphics[scale=0.45]{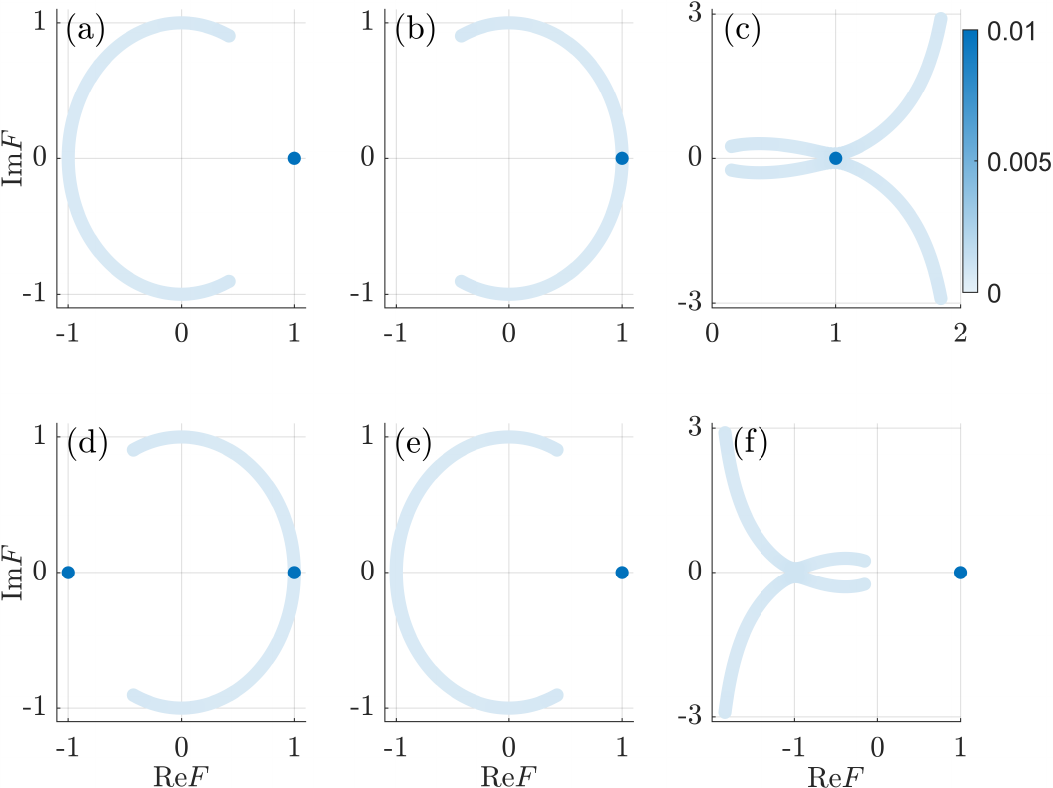}
		\par\end{centering}
	\caption{Spectra of PQNHSSH $\alpha$-chain at criticality
		with $(\alpha,\alpha')=(1,2)$, shown in terms of the Floquet operator's
		eigenvalue $F\equiv e^{-iE}$. The color of each point records the
		inverse participation ratio (IPR) \cite{NoteIPR} of related state in a lattice
		with $1000$ unit cells. System parameters are $J_{1}^{{\rm L}}=J_{1}^{{\rm R}}=J_{1}$,
		$(J_{2}^{{\rm L}},J_{2}^{{\rm R}})=(J-\gamma,J+\gamma)$, $J=3\pi/4$
		for all panels, and $J_{1}=0.32\pi$ ($0.68\pi$) for (a)--(c) ((d)--(f)).
		$\gamma=\sqrt{J^{2}-(\pi-J_{1})^{2}}$
		for (a,e), $\sqrt{J^{2}-J_{1}^{2}}$ for (b,d), $\sqrt{J^{2}+{\rm acosh}^{2}(1/\cos J_{1})}$
		for (c) and $\sqrt{J^{2}+{\rm acosh}^{2}(-1/\cos J_{1})}$ for (f).
		\label{fig:CQS}}
\end{figure}

\emph{Topology and bulk-edge correspondence}.--Under periodic boundary
condition, the Floquet operators in Eq.~(\ref{eq:U12main})
can be expressed in momentum space, and the FGBZ is
obtained as (see Sec.~II of \cite{SM} for derivations)
\begin{equation}
	\beta=(J_{\alpha'}^{{\rm R}}/J_{\alpha'}^{{\rm L}})^{1/(2\alpha'-2\alpha)}e^{i\theta},\qquad\theta\in[-\pi,\pi),\label{eq:FGBZmain}
\end{equation}
which is a circle of radius $r=|\beta|$. It
deviates from the unit circle once $|J_{\alpha'}^{{\rm L}}|\neq|J_{\alpha'}^{{\rm R}}|$,
yielding physics unique to non-Hermitian systems. The gap
closing condition along FGBZ generates
phase boundaries. Working it out explicitly with
$(J_{\alpha'}^{{\rm L}},J_{\alpha'}^{{\rm R}})\equiv(J-\gamma,J+\gamma)$,
we find the critical lines $J_{\alpha}\pm\sqrt{J^{2}-\gamma^{2}}=\nu\pi$
for $|\gamma|\leq|J|$ and $\cos(J_{\alpha})\cosh(\sqrt{\gamma^{2}-J^{2}})=(-1)^{\nu}$
for $|\gamma|>|J|$ ($\nu\in\mathbb{Z}$). They yield the curves in
Fig.~\ref{fig:PD}. They are distinct from
those found in Hermitian \cite{ZhouFgSPT1} or non-driven \cite{ZhouNHgSPT} limits, and
can be topologically allocated into four groups.

To unveil the bulk topology, we compute the
$(w_{0},w_{\pi})$ in Eq.~(\ref{eq:w0Pmain}) analytically
and report them for each phase in Fig.~\ref{fig:PD}
(see Sec.~II of \cite{SM} for details). Although the gapped
phases with $(w_{0},w_{\pi})=(\alpha,0)$ and $(\alpha',0)$ are realizable
in equilibrium \cite{ZhouNHgSPT}, the phases with $w_{\pi}=\alpha'-\alpha$
are unique to non-Hermitian Floquet setups, with each of them having
$2(\alpha'-\alpha)$ topological edge modes at $E=\pi$
[Eq.~(\ref{eq:BBCmain})]. The topology of these anomalous
Floquet phases are thus captured by our theory, even though
they both subject to NHSEs and the violation of standard bulk-edge
correspondence if $\gamma\neq0$.

\begin{figure}
	\begin{centering}
		\includegraphics[scale=0.44]{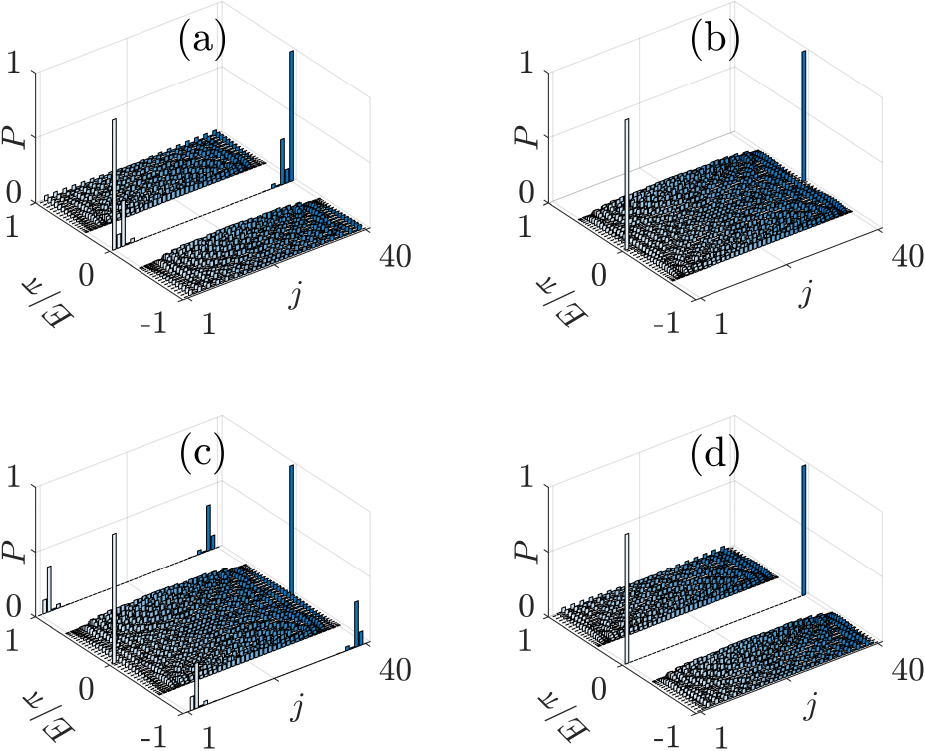}
		\par\end{centering}
	\caption{Quasienergy-resolved probability distributions $P$ of
		bulk and edge modes in the PQNHSSH $\alpha$-chain
		under OBC with $20$ unit cells. $j$ is the lattice index. Other
		parameters of (a)--(d) are the same as those of
		(a), (b), (d) and (e) in Fig.~\ref{fig:CQS}. \label{fig:CES}}
\end{figure}

Strikingly, the invariants in Eq.~(\ref{eq:w0Pmain})
are also well-defined and quantized at phase
transitions. Along the solid
critical lines where $|\tan(\sqrt{J^{2}-\gamma^{2}}/2)|=|\tan(J_{\alpha}/2)|$
in Fig.~\ref{fig:PD}, the gap closes at $E=0$ and
$(w_{0},w_{\pi})=(\alpha,0)$. They are thus topologically
nontrivial, having $N_{0}=2\alpha$ edge $0$ modes
with a gapless bulk at $E=0$ for $\alpha>0$. Along
the dash-dotted critical lines where $|\tan(\sqrt{J^{2}-\gamma^{2}}/2)\tan(J_{\alpha}/2)|=1$
in Fig.~\ref{fig:PD}, the gap closes at $E=\pi$ and
$(w_{0},w_{\pi})=(\alpha',0)$. They are always
topologically nontrivial, having $N_{0}=2\alpha'$ edge $0$
modes with a gapless bulk at $E=\pi$. The
dotted critical lines in Fig.~\ref{fig:PD} also satisfy $|\tan(\sqrt{J^{2}-\gamma^{2}}/2)\tan(J_{\alpha}/2)|=1$,
where the gap closes at $E=\pi$ and $(w_{0},w_{\pi})=(\alpha,0)$.
They are topologically nontrivial, with $2\alpha$
edge $0$ modes coexisting with a gapless bulk at
$E=\pi$ if $\alpha>0$. Notably, these two groups of
transitions are absent when the driving is switched off \cite{ZhouNHgSPT},
which implies their Floquet origins. Finally, the dashed critical lines
in Fig.~\ref{fig:PD} satisfy $|\tan(\sqrt{J^{2}-\gamma^{2}}/2)|=|\tan(J_{\alpha}/2)|$
with the gap closing at $E=0$ and $(w_{0},w_{\pi})=(\alpha,\alpha'-\alpha)$.
They are topologically nontrivial
with $2\alpha$ and $2(\alpha'-\alpha)$ edge modes at 
$E=0$ and $\pi$. Non-Hermitian Floquet gSPTs
arise along these critical lines for any $0\leq\alpha<\alpha'$.
A summary of the topology and edge modes at four distinct phase
boundaries in Fig.~\ref{fig:PD} is given in Table \ref{tab:PB}. It deserves to mention that all critical lines in
Fig.~\ref{fig:PD} refer to gap-closings between Floquet
non-Bloch bands over FGBZ. A naive application of
Bloch band theory cannot yield consistent phase boundaries and bulk-edge
correspondence at criticality.

\begin{figure}
	\begin{centering}
		\includegraphics[scale=0.46]{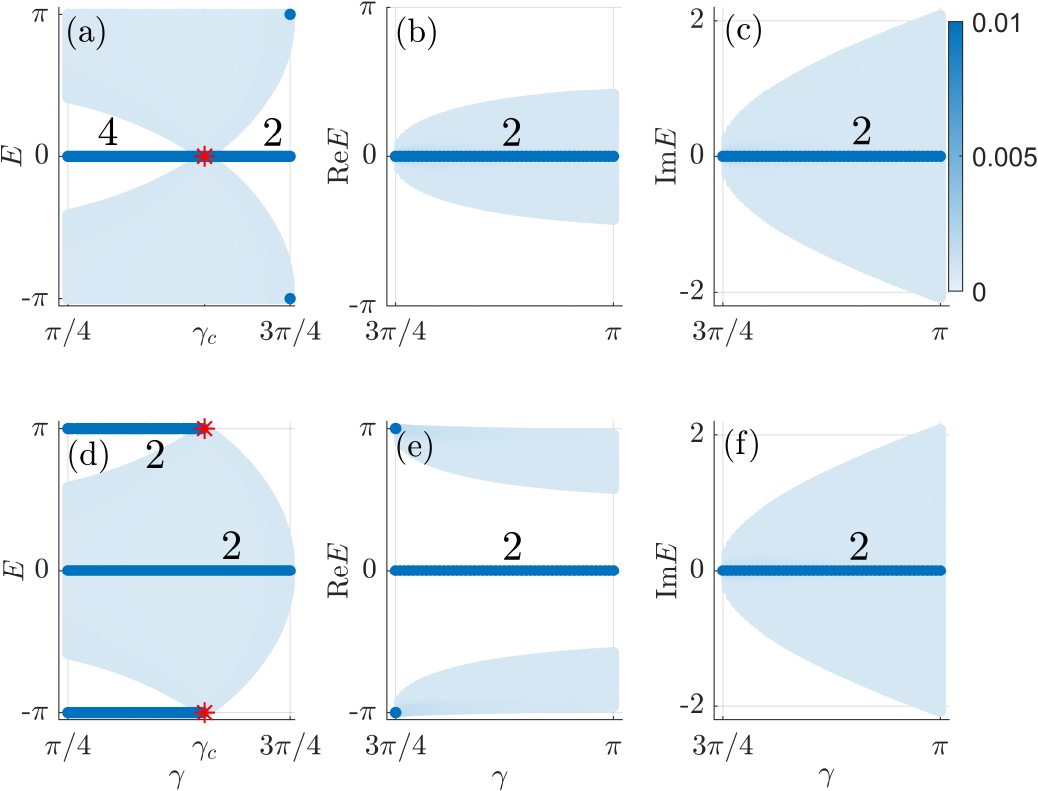}
		\par\end{centering}
	\caption{Floquet spectra along critical lines. The color of each point records the
		IPR \cite{NoteIPR} of related state in a lattice
		with $1000$ unit cells. Numbers of $0$ and $\pi$
		edge modes are given in each panel. In (a) ((d)),
		parameters are varied along the dash-dotted to dotted (dashed to solid) lines in 
		Fig.~\ref{fig:PD} for $\gamma=\pi/4\rightarrow 3\pi/4$. In
		(b)--(c) ((e)--(f)), parameters are varied along the dotted (solid) line in
		Fig.~\ref{fig:PD} for $\gamma=3\pi/4\rightarrow\pi$. \label{fig:CPT}}
\end{figure}

\emph{Critical edge modes}.--To further
reveal edge modes associated with non-Hermitian Floquet
topology at phase transitions, we obtain the quasienergies
under open boundary condition (OBC) with
$(\alpha,\alpha')=(1,2)$.
The spectra in Figs.~\ref{fig:CQS}(a), (b), (d) and (e) are obtained
at exemplary points along the dashed-dotted, solid, dashed
and dotted critical lines of Fig.~\ref{fig:PD} in its ${\cal PT}$-unbroken
region ($|\gamma|<|J|$). The probability distributions of their
edge modes are shown in (a), (b), (c) and (d) of Fig.~\ref{fig:CES}, with
the bands touching at $E=\pi$ ($F=-1$), $0$
($F=1$), $0$ and $\pi$. Our bulk
theory {[}Eq.~(\ref{eq:BBCmain}){]} predicts the numbers of
edge modes $(N_{0},N_{\pi})=(4,0)$, $(2,0)$,
$(2,2)$ and $(2,0)$ at these critical points, which are in perfect
agreement with those observed in Fig.~\ref{fig:CES}, verifying
the bulk-edge correspondence at phase transitions. In the ${\cal PT}$-broken
region ($|\gamma|>|J|$),
the range of exponential $e^{-iE}$ goes beyond the unit circle
$|F|=1$, as shown in Figs.~\ref{fig:CQS}(c) and \ref{fig:CQS}(f). Nevertheless,
we find two edge $0$ modes at both these critical
points, regardless of whether the bulk gap closes at $E=0$
{[}$F=1$ in Fig.~\ref{fig:CQS}(c){]} or $E=\pi$ {[}$F=-1$ in 
Fig.~\ref{fig:CQS}(f){]}, and their distributions are identical
to those in (b) and (d) of Fig.~\ref{fig:CES}. Lying along
the solid and dotted lines in Fig.~\ref{fig:PD}, the numbers
of edge modes at these critical points are $(N_{0},N_{\pi})=(2,0)$,
verifying again the Eq.~(\ref{eq:BBCmain}).
Overall, we confirmed the nontrivial topology and edge modes
surviving criticality in our setting, regardless
of whether the underlying quasienergy spectrum is in ${\cal PT}$-invariant
or ${\cal PT}$-broken regime.

\begin{figure}
	\begin{centering}
		\includegraphics[scale=0.43]{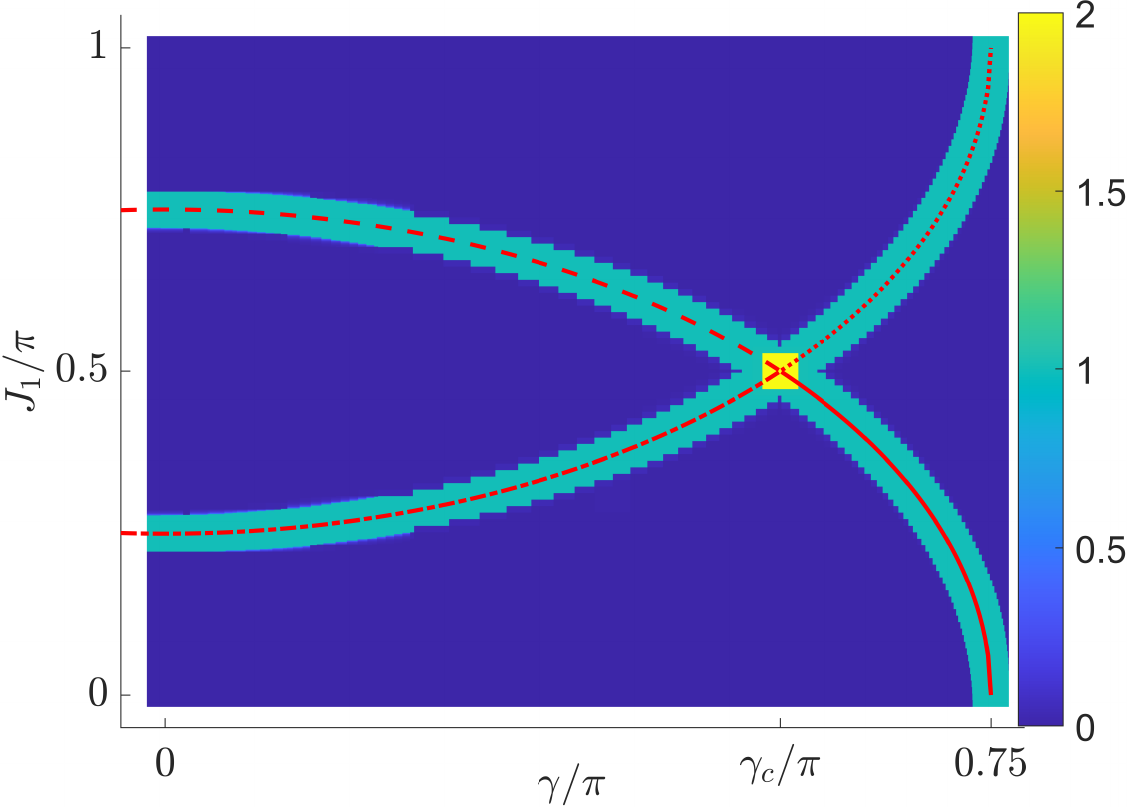}
		\par\end{centering}
	\caption{Central charge of PQNHSSH $\alpha$-chain with $(\alpha,\alpha')=(1,2)$.
		System parameters are $(J_{2}^{{\rm L}},J_{2}^{{\rm R}})=(J-\gamma,J+\gamma)$
		with $J=3\pi/4$. The length of lattice is $L=1000$.\label{fig:CC}}
\end{figure}

For the case with $(\alpha,\alpha')=(1,2)$, we
find two Floquet eigenstates
at quasienergy $E=0$ and one
at $E=\pi$ when considering a half-infinite chain with
the OBC taken at its left edge. Analytically,
these eigenmodes are given by (see Sec.~II of \cite{SM} for details)
$|\varphi_{0,1}\rangle= \hat{a}_{1}^{\dagger}|\emptyset\rangle$,
$|\varphi_{0,2}\rangle= \sum_{n=1}^{\infty}\left[-\frac{\tan(J_{1}/2)}{\tan(J_{2}/2)}\right]^{n}[\cos(J_{1}/2)\hat{a}_{n+1}^{\dagger}-i\sin(J_{1}/2)\hat{b}_{n}^{\dagger}]|\emptyset\rangle$, and $|\varphi_{\pi}\rangle= \sum_{n=1}^{\infty}\left[\frac{1}{\tan(J_{1}/2)\tan(J_{2}/2)}\right]^{n}[\sin(J_{1}/2)\hat{a}_{n+1}^{\dagger}+i\cos(J_{1}/2)\hat{b}_{n}^{\dagger}]|\emptyset\rangle$,
where $J_{2}\equiv\sqrt{J^{2}-\gamma^{2}}$. Each of the states, when
exists, has a degenerate partner in thermodynamic limit due to the
sublattice symmetry. We see that the state $|\varphi_{0,1}\rangle$
represents an edge $0$ mode throughout the phase diagram,
ensuring the nontrivial topology of all critical lines. The mode $|\varphi_{0,2}\rangle$
is localized, delocalized and eliminated when $|\tan(J_{1}/2)/\tan(J_{2}/2)|<1$,
$=1$ and $>1$, respectively. It thus depicts an edge $0$ mode when
$|\tan(J_{2}/2)|>|\tan(J_{1}/2)|$, which survives along the critical
lines where $|\tan(J_{1}/2)\tan(J_{2}/2)|=1$ with the bulk gap closing
at quasienergy $\pi$. The mode $|\varphi_{\pi}\rangle$ is localized,
delocalized and eliminated when $|\tan(J_{1}/2)\tan(J_{2}/2)|>1$,
$=1$ and $<1$, respectively. It then describes an edge $\pi$ mode
when $|\tan(J_{1}/2)\tan(J_{2}/2)|>1$, which survives along the critical
lines where $|\tan(J_{1}/2)|=|\tan(J_{2}/2)|$ and the bulk gap closes
at quasienergy $0$. These results are consistent with the
predictions collected in Table \ref{tab:PB}.

\begin{table*}
	\caption{Summary of the topology and edge modes surviving criticality in PQNHSSH
		$\alpha$-chain along phase boundaries in Fig.~\ref{fig:PD}.	
		Here, $J_{\alpha'}\equiv\sqrt{J_{\alpha'}^{{\rm L}}J_{\alpha'}^{{\rm R}}}$
		and $J_{\alpha'}^{{\rm L}}J_{\alpha'}^{{\rm R}}=J^{2}-\gamma^{2}$.
		Two Floquet bands are touched at the
		gapless quasienergy.
		The case of having $\pi$ edge modes when the bulk spectrum is gapless at $E=\pi$ is discussed in Sec.~II.H of \cite{SM}.
		\label{tab:PB}}
	\centering{}%
	\begin{tabular}{|c|c|c|c|c|}
		\hline 
		\multirow{2}{*}{Critical line} & \multirow{2}{*}{Condition} & Gapless & Topological & Number of edge\tabularnewline
		&  & \multirow{1}{*}{quasienergy} & invariants $(w_{0},w_{\pi})$ & modes $(N_{0},N_{\pi})$\tabularnewline
		\hline 
		\hline 
		\multirow{2}{*}{Solid} & $|\tan(J_{\alpha'}/2)|=|\tan(J_{\alpha}/2)|$ & \multirow{2}{*}{$E=0$} & \multirow{2}{*}{$(\alpha,0)$} & \multirow{2}{*}{$(2\alpha,0)$}\tabularnewline
		& $|\tan(J_{\alpha'}/2)\tan(J_{\alpha}/2)|<1$ &  &  & \tabularnewline
		\hline 
		\multirow{2}{*}{Dashed} & $|\tan(J_{\alpha'}/2)|=|\tan(J_{\alpha}/2)|$ & \multirow{2}{*}{$E=0$} & \multirow{2}{*}{$(\alpha,\alpha'-\alpha)$} & \multirow{2}{*}{$(2\alpha,2\alpha'-2\alpha)$}\tabularnewline
		& $|\tan(J_{\alpha'}/2)\tan(J_{\alpha}/2)|>1$ &  &  & \tabularnewline
		\hline 
		\multirow{2}{*}{Dotted} & $|\tan(J_{\alpha'}/2)\tan(J_{\alpha}/2)|=1$ & \multirow{2}{*}{$E=\pm\pi$} & \multirow{2}{*}{$(\alpha,0)$} & \multirow{2}{*}{$(2\alpha,0)$}\tabularnewline
		& $|\tan(J_{\alpha'}/2)|<|\tan(J_{\alpha}/2)|$ &  &  & \tabularnewline
		\hline 
		\multirow{2}{*}{Dash-dotted} & $|\tan(J_{\alpha'}/2)\tan(J_{\alpha}/2)|=1$ & \multirow{2}{*}{$E=\pm\pi$} & \multirow{2}{*}{$(\alpha',0)$} & \multirow{2}{*}{$(2\alpha',0)$}\tabularnewline
		& $|\tan(J_{\alpha'}/2)|>|\tan(J_{\alpha}/2)|$ &  &  & \tabularnewline
		\hline 
		\multirow{2}{*}{Multicritical} & $|\tan(J_{\alpha'}/2)|=|\tan(J_{\alpha}/2)|$ & \multirow{2}{*}{$E=0,\pm\pi$} & \multirow{2}{*}{$(\alpha,0)$} & \multirow{2}{*}{$(2\alpha,0)$}\tabularnewline
		& $|\tan(J_{\alpha'}/2)\tan(J_{\alpha}/2)|=1$ &  &  & \tabularnewline
		\hline 
	\end{tabular}
\end{table*}

\emph{Phase transitions between gSPTs}.--With four critical lines of
different topology, we expect to see phase transitions
along topological phase boundaries via changing parameters
across the multicritical point.
In Fig.~\ref{fig:CPT}, we present the spectrum of our model
with $(\alpha,\alpha')=(1,2)$ along phase boundaries, where the
star denotes the multicritical point $(J_{1},\gamma_{c})=(\pi/2,\sqrt{5}\pi/4)$.
In (a) and (d) of Fig.~\ref{fig:CPT}, we find a change of $0$
and $\pi$ edge modes by $2$ when going from one to the other side
of multicritical point, during which the spectrum remains
gapless at $E=\pi$ and $0$, respectively. The $(w_{0},w_{\pi})$ in Eq.~(\ref{eq:w0Pmain}) change from
$(2,0)$ {[}$(1,1)${]} to $(1,0)$ across the multicritical point
in Fig.~\ref{fig:CPT}(a) {[}\ref{fig:CPT}(d){]}. These cases
represent two examples of phase transitions between distinct
non-Hermitian Floquet gSPTs, which are identified by the
survival of edge zero modes at criticality. In Figs.~\ref{fig:CPT}(b,c)
and \ref{fig:CPT}(e,f), we show the spectra along
two critical lines in the ${\cal PT}$-broken region, with bulk
bands touched at $E=0$ and $\pi$. There are no multicritical
points on these phase boundaries, and we find two
edge $0$ modes in both cases. In all examples, the degeneracy of
$0$ and $\pi$ edge modes are protected by the sublattice
symmetry.

The necessity of discriminating phase transitions via
topology can also be seen from entanglement viewpoints.
In Fig.~\ref{fig:CC}, we extract the central charge $c$ \cite{EECFT,LiHaldane,ESEE01,ESEE02,ESEE03,ESEE04} of bipartite
EE $S(L,l)$ at half-filling in a ${\cal PT}$-unbroken region ($|\gamma|<|J|,J_{\alpha}>0$).
A finite-size scaling is performed vs the fitting curve $S(L,l)\sim c\ln[\sin(\pi l/L)]/3$
for EE under a fixed system size $L$ and varied subsystem sizes $l$ to exact $c$ (see Sec.~III of \cite{SM} for details).
Despite the
multicritical point where $c\simeq2$, we find $c\simeq1$ along
critical lines. The former comes about due to the simultaneous gap
closing of Floquet non-Bloch bands at $E=0$ and $\pi$,
which is thus of nonequilibrium origin. The latter implies that the
critical lines in Fig.~\ref{fig:CC} may belong to the same universality
class, which can only be distinguished by
their topology and edge modes. For other
hopping ranges $(\alpha,\alpha')$, we expect $c=\alpha'-\alpha$
along critical lines (with $J_{1}\mapsto J_{\alpha}$
in Fig.~\ref{fig:CC}) and $c=2\alpha'-2\alpha$ at multicritical
points.

\emph{Discussion}.--Experimentally, the Hermitian version of
our model is realizable in acoustics \cite{gSPTExp03}, where
the quasienergies are accessible by a pump-probe
method, and the driving is emulated by
modulations along the propagation direction of sound.
By recording both the amplitude and phase of pressure fields, the
Floquet bands can be constructed from phase-resolved measurements,
allowing the identification of whether the system is away from
or at a gapless phase transition \cite{gSPTExp03}.
The critical $0$ and $\pi$ modes can be further detected by tracking the
dynamics of edge excitations \cite{ChengPRL2022}, in which the former 
shows a periodic-revival while the latter shows a periodic-doubling
(see Sec.~III of \cite{SM} for more details).
Finally, the non-Hermiticity,
as introduced by asymmetric couplings in our case, can be realized
with directional amplifiers in acoustic crystals \cite{ZhangNC2021}.
Acoustic metamaterials thus offer a flexible means to realize
our model and detect its topology and edge modes
surviving criticality. Other setups where our model
is within reach include photonics \cite{NHQW01,NHQW02,NHQW03}
and electrical circuits \cite{NHEC01,NHEC02,NHEC03}. 

Overall, we introduced a unique class of driven non-Hermitian models that are exactly solvable and unveiled a distinctive type of nonequilibrium gapless topology. A theoretical framework, based on a unified description of bulk Floquet non-Bloch bands, edge states, bulk-edge correspondence and entanglement was developed to characterize the topology and edge modes surviving criticality. Moreover, we identified $\pi$ edge modes sandwiching between gapless Floquet non-Bloch bands and critical topological transitions via closing Floquet non-Bloch band gaps at $E=\pi$, with both being unavailable in static or Hermitian setups and are thus of non-Hermitian Floquet origins. These discoveries refined the organization of phase transitions in driven open systems, in which topology enriches criticality beyond the standard paradigms of continuous and topological phase transitions.
In future work, it would
be interesting to generalize our scheme to systems with multiple
bands \cite{NAFTI01,NAFTI02,NAFTI03}, in other symmetry classes
\cite{GongPRX2018}, and beyond 1D cases
\cite{AmoebaPRX2024}. Exceptional topology \cite{ETRMP} of non-Hermitian
Floquet origin and its impact on critical edge modes
are awaited to be further clarified. The roles of disorder
and interactions \cite{MB1,MB2,MB3,MB4} in non-Hermitian Floquet gSPTs
also deserve more thorough explorations.

\emph{Acknowledgments}.--L.Z. is supported by the NSFC (Grants No.~12275260 and No.~11905211), the Fundamental Research Funds for the Central Universities (Grant No.~202364008), and the Young Talents Project of Ocean University of China.



\begin{thebibliography}{99}

\bibitem{gSPT01} T. Scaffidi, D. E. Parker, and R. Vasseur, Gapless
Symmetry-Protected Topological Order, Phys. Rev. X \textbf{7}, 041048
(2017).

\bibitem{gSPT02} R. Verresen, R. Thorngren, N. G. Jones, and F. Pollmann,
Gapless Topological Phases and Symmetry-Enriched Quantum Criticality,
Phys. Rev. X \textbf{11}, 041059 (2021).

\bibitem{gSPT03} Y. Baum, T. Posske, I. C. Fulga, B. Trauzettel,
and A. Stern, Coexisting Edge States and Gapless Bulk in Topological
States of Matter, Phys. Rev. Lett. \textbf{114}, 136801 (2015).

\bibitem{gSPT04} A. Keselman and E. Berg, Gapless symmetry-protected
topological phase of fermions in one dimension, Phys. Rev. B \textbf{91},
235309 (2015).

\bibitem{gSPT05} S. C. Furuya and M. Oshikawa, Symmetry protection
of critical phases and a global anomaly in $1+1$ dimensions, Phys.
Rev. Lett. \textbf{118}, 021601 (2017).

\bibitem{gSPT06} R. Verresen, R. Moessner, and F. Pollmann, One-dimensional
symmetry protected topological phases and their transitions, Phys.
Rev. B \textbf{96}, 165124 (2017).

\bibitem{gSPT07} R. Verresen, N. G. Jones, and F. Pollmann, Topology
and Edge Modes in Quantum Critical Chains, Phys. Rev. Lett. \textbf{120},
057001 (2018).

\bibitem{gSPT08} W. Berdanier, M. Kolodrubetz, S. A. Parameswaran,
and R. Vasseur, Floquet quantum criticality, Proc. Natl. Acad. Sci.
U.S.A. \textbf{115}, 9491-9496 (2018).

\bibitem{gSPT09} N. G. Jones and R. Verresen, Asymptotic Correlations
in Gapped and Critical Topological Phases of 1D Quantum Systems, J.
Stat. Phys. \textbf{175}, 1164--1213 (2019).

\bibitem{gSPT10} D. E. Parker, R. Vasseur and T. Scaffidi, Topologically
protected long edge coherence times in symmetry-broken phases, Phys.
Rev. Lett. \textbf{122}, 240605 (2019).

\bibitem{gSPT11} R. Verresen, Topology and edge states survive quantum
criticality between topological insulators, arXiv:2003.05453.

\bibitem{gSPT12} O. Balabanov, D. Erkensten, and H. Johannesson,
Topology of critical chiral phases: Multiband insulators and superconductors,
Phys. Rev. Res. \textbf{3}, 043048 (2021).

\bibitem{gSPT13} R. Thorngren, A. Vishwanath and R. Verresen, Intrinsically
gapless topological phases, Phys. Rev. B \textbf{104}, 075132 (2021).

\bibitem{gSPT14} C. M. Duque, H.-Y. Hu, Y.-Z. You, V. Khemani, R.
Verresen, and R. Vasseur, Topological and symmetry-enriched random
quantum critical points, Phys. Rev. B \textbf{103}, L100207 (2021).

\bibitem{gSPT15} U. Borla, R. Verresen, J. Shah and S. Moroz, Gauging
the Kitaev chain, SciPost Phys. \textbf{10}, 148 (2021).

\bibitem{gSPT16} R. R. Kumar, Y. R. Kartik, S. Rahul, and S. Sarkar,
Multi-critical topological transition at quantum criticality, Sci.
Rep. \textbf{11}, 1004 (2021).

\bibitem{gSPT17} X.-J. Yu, R.-Z. Huang, H.-H. Song, L. Xu, C. Ding
and L. Zhang, Conformal boundary conditions of symmetry-enriched quantum
critical spin chains, Phys. Rev. Lett. \textbf{129}, 210601 (2022).

\bibitem{gSPT18} Y. Hidaka, S. C. Furuya, A. Ueda and Y. Tada, Gapless
symmetry-protected topological phase of quantum antiferromagnets on
anisotropic triangular strip, Phys. Rev. B \textbf{106}, 144436 (2022).

\bibitem{gSPT19} R. Ma, L. Zou and C. Wang, Edge physics at the deconfined
transition between a quantum spin Hall insulator and a superconductor,
SciPost Phys. \textbf{12}, 196 (2022).

\bibitem{gSPT20} N. G. Jones, R. Thorngren, and R. Verresen, Bulk-Boundary
Correspondence and Singularity-Filling in Long-Range Free-Fermion
Chains, Phys. Rev. Lett. \textbf{130}, 246601 (2023).

\bibitem{gSPT21} R. Wen and A. C. Potter, Bulk-boundary correspondence
for intrinsically gapless symmetry-protected topological phases from
group cohomology, Phys. Rev. B \textbf{107}, 245127 (2023).

\bibitem{gSPT22} H. Yang, L. Li, K. Okunishi and H. Katsura, Duality,
criticality, anomaly, and topology in quantum spin-$1$ chains, Phys.
Rev. B \textbf{107}, 125158 (2023).

\bibitem{gSPT23} R. R. Kumar, N. Roy, Y. R. Kartik, S. Rahul, and
S. Sarkar, Signatures of topological phase transition on a quantum
critical line, Phys. Rev. B \textbf{107}, 205114 (2023).

\bibitem{gSPT24} S. Mondal, A. Agarwala, T. Mishra, and A. Prakash,
Symmetry-enriched criticality in a coupled spin ladder, Phys. Rev.
B \textbf{108}, 245135 (2023).

\bibitem{gSPT25} S. Prembabu, R. Thorngren, and R. Verresen, Boundary-deconfined
quantum criticality at transitions between symmetry-protected topological
chains, Phys. Rev. B \textbf{109}, L201112 (2024).

\bibitem{gSPT26} L. Li, M. Oshikawa and Y. Zheng, Decorated defect
construction of gapless-SPT states, SciPost Phys. \textbf{17}, 013
(2024).

\bibitem{gSPT27} L. Li, M. Oshikawa and Y. Zheng, Intrinsically/purely
gapless-SPT from non-invertible duality transformations, SciPost Phys.
\textbf{18}, 153 (2025).

\bibitem{gSPT28} R. Flores-Calder\'on, E. J. K\"onig, and A. M.
Cook, Topological Quantum Criticality from Multiplicative Topological
Phases, Phys. Rev. Lett. \textbf{134}, 116602 (2025).

\bibitem{gSPT29} H. Jia, J. Hu, R.-Y. Zhang, Y. Xiao, D. Wang, M.
Wang, S. Ma, X. Ouyang, Y. Zhu, and C. T. Chan, Unconventional Topological
Edge States In One-Dimensional Non-Hermitian Gapless Systems Stemming
from Nonisolated Hypersurface Singularities, Phys. Rev. Lett. \textbf{134},
206603 (2025).

\bibitem{gSPT30} C. Song, Zero Curvature Condition for Quantum Criticality,
Phys. Rev. Lett. \textbf{134}, 240202 (2025).

\bibitem{gSPT31} X. Shen, Z. Wu, and S.-K. Jian, Boundary and defect
criticality in topological insulators and superconductors, Phys. Rev.
B \textbf{112}, L041118 (2025).

\bibitem{gSPT32} A. Chatterjee, W. Ji, and X.-G. Wen, Emergent generalized
symmetry and maximal symmetry topological order, Phys. Rev. B \textbf{112},
115142 (2025).

\bibitem{gSPT33} G. Cardoso, H.-C. Yeh, L. Korneev, A. G. Abanov,
and A. Mitra, Gapless Floquet topology, Phys. Rev. B \textbf{111},
125162 (2025).

\bibitem{gSPT34} R. Wen and A. C. Potter, Classification of $1+1$D
gapless symmetry protected phases via topological holography, Phys.
Rev. B \textbf{111}, 115161 (2025).

\bibitem{gSPT35} X. Zhou, S. Jia, and J.-S. Pan, Interaction-induced
phase transitions at topological quantum criticality of an extended
Su-Schrieffer-Heeger model, Phys. Rev. B \textbf{111}, 195117 (2025).

\bibitem{gSPT36} L. Li, R.-Z. Huang, and W. Cao, Noninvertible symmetry-enriched
quantum critical point, Phys. Rev. B \textbf{112}, L081113 (2025).

\bibitem{gSPT37} S.-J. Huang and M. Cheng, Topological holography,
quantum criticality, and boundary states, SciPost Phys. \textbf{18},
213 (2025).

\bibitem{gSPT38} R. Barad, Q. Tang, W. Zhu, and X. Wen, Universal
time evolution of string order parameter in quantum critical systems
with boundary invertible or noninvertible symmetry breaking, Phys.
Rev. B \textbf{111}, 165121 (2025).

\bibitem{gSPTExp01} R. Shen, T. Chen, B. Yang, Y. Zhong, and C. H.
Lee, Robust simulations of many-body symmetry-protected topological
phase transitions on a quantum processor, npj Quantum Inf. \textbf{11},
179 (2025).

\bibitem{gSPTExp02} Z. Tan, K. Wang, S. Yang, et. al., Exploring
nontrivial topology at quantum criticality in a superconducting processor,
arXiv:2501.04679.

\bibitem{gSPTExp03} Z. Cheng, X. Zhang, X.-J. Yu, L. Zhou, J. Gong,
and B. Zhang, Observation of critical topological phase transition,
\emph{Nature, under review}.

\bibitem{FTP01} T. Oka and H. Aoki, Photovoltaic Hall effect in graphene,
Phys. Rev. B \textbf{79}, 081406(R) (2009).

\bibitem{FTP02} G. Jotzu, M. Messer, R. Desbuquois, M. Lebrat, T.
Uehlinger, D. Greif, and T. Esslinger, Experimental realization of
the topological Haldane model with ultracold fermions, Nature \textbf{515},
237--240 (2014).

\bibitem{FTP03} J. W. McIver, B. Schulte, F.-U. Stein, T. Matsuyama,
G. Jotzu, G. Meier, and A. Cavalleri, Light-induced anomalous Hall
effect in graphene, Nat. Phys. \textbf{16}, 38--41 (2020).

\bibitem{FTP04} D. Y. H. Ho and J. Gong, Quantized Adiabatic Transport
In Momentum Space, Phys. Rev. Lett. \textbf{109}, 010601 (2012).

\bibitem{FTP05} Q.-J. Tong, J.-H. An, J. Gong, H.-G. Luo, and C.
H. Oh, Generating many Majorana modes via periodic driving: A superconductor
model, Phys. Rev. B \textbf{87}, 201109(R) (2013).

\bibitem{FTP06} K. Yang, S. Xu, L. Zhou, Z. Zhao, T. Xie, Z. Ding,
W. Ma, J. Gong, F. Shi, and J. Du, Observation of Floquet topological
phases with large Chern numbers, Phys. Rev. B \textbf{106}, 184106
(2022).

\bibitem{FTP07} L. Jiang, T. Kitagawa, J. Alicea, A. R. Akhmerov,
D. Pekker, G. Refael, J. I. Cirac, E. Demler, M. D. Lukin, and P.
Zoller, Majorana Fermions in Equilibrium and in Driven Cold-Atom Quantum
Wires, Phys. Rev. Lett. \textbf{106}, 220402 (2011).

\bibitem{FTP08} M. S. Rudner, N. H. Lindner, E. Berg, and M. Levin,
Anomalous Edge States and the Bulk-Edge Correspondence for Periodically
Driven Two-Dimensional Systems, Phys. Rev. X \textbf{3}, 031005 (2013).

\bibitem{FTP09} K. Wintersperger, C. Braun, F. N. \"Unal, A. Eckardt,
M. D. Liberto, N. Goldman, I. Bloch, and M. Aidelsburger, Realization
of an anomalous Floquet topological system with ultracold atoms, Nat.
Phys. \textbf{16}, 1058--1063 (2020).

\bibitem{BiorthogonalQM} D. C. Brody, Biorthogonal quantum mechanics,
J. Phys. A: Math. Theor. \textbf{47}, 035305 (2014).

\bibitem{NHTP01} K. Kawabata, K. Shiozaki, M. Ueda, and M. Sato,
Symmetry and Topology in Non-Hermitian Physics, Phys. Rev. X \textbf{9},
041015 (2019).

\bibitem{NHTP02} H. Zhou and J. Y. Lee, Periodic table for topological
bands with non-Hermitian symmetries, Phys. Rev. B \textbf{99}, 235112
(2019).

\bibitem{NHTP03} C. M. Bender and D. W. Hook, ${\cal PT}$-symmetric
quantum mechanics, Rev. Mod. Phys. \textbf{96}, 045002 (2024).

\bibitem{EP01} W. D. Heiss, The physics of exceptional points, J.
Phys. A: Math. Theor. \textbf{45}, 444016 (2012).

\bibitem{EP02} J. Wiersig, Enhancing the sensitivity of frequency
and energy splitting detection by using exceptional points: Application
to microcavity sensors for single-particle detection, Phys. Rev. Lett.
\textbf{112}, 203901 (2014).

\bibitem{EP03} M.-A. Miri and A. Al\`u, Exceptional points in optics
and photonics, Science \textbf{363}, eaar7709 (2019).

\bibitem{NHSE01} S. Yao and Z. Wang, Edge States and Topological
Invariants of Non-Hermitian Systems, Phys. Rev. Lett. \textbf{121},
086803 (2018).

\bibitem{NHSE02} F. K. Kunst, E. Edvardsson, J. C. Budich, and E.
J. Bergholtz, Biorthogonal Bulk-Boundary Correspondence in Non-Hermitian
Systems, Phys. Rev. Lett. \textbf{121}, 026808 (2018).

\bibitem{NHSE03} V. M. Martinez Alvarez, J. E. Barrios Vargas, and
L. E. F. Foa Torres, Non-Hermitian robust edge states in one dimension:
Anomalous localization and eigenspace condensation at exceptional
points, Phys. Rev. B \textbf{97}, 121401(R) (2018).

\bibitem{NHFTPrev} L. Zhou and D.-J. Zhang, Non-Hermitian Floquet
Topological Matter---A Review, Entropy \textbf{25}, 1401 (2023).

\bibitem{ZhouFgSPT1} L. Zhou, J. Gong, and X.-J. Yu, Topological
edge states at Floquet quantum criticality, Commun. Phys. \textbf{8},
214 (2025).

\bibitem{ZhouFgSPT2} L. Zhou, R. Wang, and J. Pan, Gapless higher-order
topology and corner states in Floquet systems, Phys. Rev. Res. \textbf{7},
023079 (2025).

\bibitem{ZhouFgSPT3} L. Zhou, F. Zhang, and J. Pan, Floquet M\"obius
topological insulators, Phys. Rev. B \textbf{112}, 134302 (2025).

\bibitem{GBZ01} K. Yokomizo and S. Murakami, Non-Bloch Band Theory
of Non-Hermitian Systems, Phys. Rev. Lett. \textbf{123}, 066404 (2019).

\bibitem{GBZ02} Z. Yang, K. Zhang, C. Fang, and J. Hu, Non-Hermitian
Bulk-Boundary Correspondence and Auxiliary Generalized Brillouin Zone
Theory, Phys. Rev. Lett. \textbf{125}, 226402 (2020).

\bibitem{GBZ03} Y. Cao, Y. Li, and X. Yang, Non-Hermitian bulk-boundary
correspondence in a periodically driven system, Phys. Rev. B \textbf{103},
075126 (2021).

\bibitem{NHFTP01} L. Zhou and J. Gong, Non-Hermitian Floquet topological
phases with arbitrarily many real-quasienergy edge states, Phys. Rev.
B \textbf{98}, 205417 (2018).

\bibitem{NHFTP02} L. Xiao, T. Deng, K. Wang, G. Zhu, Z. Wang, W.
Yi, and P. Xue, Non-Hermitian bulk--boundary correspondence in quantum
dynamics, Nat. Phys. \textbf{16}, 761--766 (2020).

\bibitem{NHFTP03} X. Zhang and J. Gong, Non-Hermitian Floquet topological
phases: Exceptional points, coalescent edge modes, and the skin effect,
Phys. Rev. B \textbf{101}, 045415 (2020).

\bibitem{FGBZ02} L. Zhou, Y. Gu, and J. Gong, Dual topological characterization of non-Hermitian Floquet phases, Phys. Rev. B {\bf 103}, L041404 (2021).

\bibitem{SM} In SM, see Sec.~I for the general theoretical framework; Sec.~II for the application of our theory to PQNHSSH $\alpha$ chain; and Sec.~III for entanglement aspects of the PQNHSSH $\alpha$-chain.

\bibitem{ZP01} C. H. Lee and R. Thomale, Anatomy of skin modes and
topology in non-Hermitian systems, Phys. Rev. B \textbf{99}, 201103(R)
(2019).

\bibitem{GBZAL2} K. Zhang, Z. Yang, and C. Fang, Correspondence between Winding Numbers and Skin Modes in Non-Hermitian Systems, Phys. Rev. Lett. {\bf 125}, 126402 (2020).

\bibitem{ZP02} S. Verma and M. J. Park, Topological phase transitions
of generalized Brillouin zone, Commun. Phys. \textbf{7}, 21 (2024).

\bibitem{ZP03} J. Zhong, H. Wang, and S. Fan, Pole and zero edge
state invariant for one-dimensional non-Hermitian sublattice symmetry,
Phys. Rev. B \textbf{110}, 214113 (2024).

\bibitem{GBZAL5} J. Zhong, H. Wang, A. N. Poddubny, and S. Fan, Topological Nature of Edge States for One-Dimensional Systems without Symmetry Protection, Phys. Rev. Lett. {\bf 135}, 016601 (2025).

\bibitem{ZhouNHgSPT} L. Zhou, R. Jing, and S. Wu, Topological characterization
of phase transitions and critical edge states in one-dimensional non-Hermitian
systems with sublattice symmetry, Front. Phys. {\bf 21}, 075202 (2026).

\bibitem{STF01} J. K. Asb\'oth and H. Obuse, Bulk-boundary correspondence
for chiral symmetric quantum walks, Phys. Rev. B \textbf{88}, 121406(R)
(2013).

\bibitem{STF02} M. Rodriguez-Vega and B. Seradjeh, Universal Fluctuations
of Floquet Topological Invariants at Low Frequencies, Phys. Rev. Lett.
\textbf{121}, 036402 (2018).

\bibitem{STF03} L. Zhou and J. Gong, Floquet topological phases in
a spin-$1/2$ double kicked rotor, Phys. Rev. A \textbf{97}, 063603
(2018).

\bibitem{NoteIPR} The IPR of a bulk state is inversely proportional to the system size $L$, and it goes to $0$ when $L\rightarrow\infty$. The IPR of an edge mode converges to a finite value when $L\rightarrow\infty$ due to its exponential localization at the edge.

\bibitem{EECFT} P. Calabrese and J. Cardy, Entanglement entropy and quantum field theory, J. Stat. Mech. {\bf 2004}, P06002 (2004).

\bibitem{LiHaldane} H. Li and F. D. M. Haldane, Entanglement Spectrum
as a Generalization of Entanglement Entropy: Identification of Topological
Order in Non-Abelian Fractional Quantum Hall Effect States, Phys.
Rev. Lett. \textbf{101}, 010504 (2008).

\bibitem{ESEE01} I. Peschel and V. Eisler, Reduced density matrices
and entanglement entropy in free lattice models, J. Phys. A: Math.
Theor. \textbf{42}, 504003 (2009).

\bibitem{ESEE02} L. Zhou, Entanglement spectrum and entropy in Floquet
topological matter, Phys. Rev. Res. \textbf{4}, 043164 (2022).

\bibitem{ESEE03} L.-M. Chen, Y. Zhou, S. A. Chen, and P. Ye, Quantum
Entanglement and Non-Hermiticity in Free-Fermion Systems, Chin. Phys.
Lett. \textbf{41}, 127302 (2024).

\bibitem{ESEE04} L. Zhou, Quantum geometry and geometric entanglement
entropy of one-dimensional Floquet topological matter, Phys. Rev.
B \textbf{110}, 054310 (2024).

\bibitem{ChengPRL2022} Z. Cheng, R. W. Bomantara, H. Xue, W. Zhu,
J. Gong, and B. Zhang, Observation of $\pi/2$ modes in an acoustic
Floquet system, Phys. Rev. Lett. \textbf{129}, 254301 (2022).

\bibitem{ZhangNC2021} L. Zhang, Y. Yang, Y. Ge, Y.-J. Guan, Q. Chen,
Q. Yan, F. Chen, R. Xi, Y. Li, D. Jia, S.-Q. Yuan, H.-X. Sun, H. Chen,
and B. Zhang, Acoustic non-Hermitian skin effect from twisted winding
topology, Nat. Commun. \textbf{12}, 6297 (2021).

\bibitem{NHQW01} L. Xiao, T. Deng, K. Wang, Z. Wang, W. Yi, and P.
Xue, Observation of Non-Bloch Parity-Time Symmetry and Exceptional
Points, Phys. Rev. Lett. \textbf{126}, 230402 (2021).

\bibitem{NHQW02} J. Zhu, Y.-L. Mao, H. Chen, K.-X. Yang, L. Li, B.
Yang, Z.-D. Li, and J. Fan, Observation of Non-Hermitian Edge Burst
Effect in One-Dimensional Photonic Quantum Walk, Phys. Rev. Lett.
\textbf{132}, 203801 (2024).

\bibitem{NHQW03} L. Xiao, W.-T. Xue, F. Song, Y.-M. Hu, W. Yi, Z.
Wang, and P. Xue, Observation of Non-Hermitian Edge Burst in Quantum
Dynamics, Phys. Rev. Lett. \textbf{133}, 070801 (2024).

\bibitem{NHEC01} T. Helbig, T. Hofmann, S. Imhof, M. Abdelghany,
T. Kiessling, L. W. Molenkamp, C. H. Lee, A. Szameit, M. Greiter,
and R. Thomale, Generalized bulk--boundary correspondence in non-Hermitian
topolectrical circuits, Nat. Phys. \textbf{16}, 747--750 (2020).

\bibitem{NHEC02} X. Zhou, W. Zhang, W. Cao, and X. Zhang, Non-Hermitian
Floquet Topological Sensors for Ultrasensitive Detection of Dynamic
Signals, Phys. Rev. Lett. \textbf{135}, 106601 (2025).

\bibitem{NHEC03} J. Zhang, W. Song, H. Li, Z. Kuang, Z. Xiong, L.
Zhou, J. Liu, and W.-L. Zhao, Observing the exponential growth of
the eigenmodes in the absence of coalescence for a non-Hermitian circuit
with an unavoidable inductor dissipation, Phys. Rev. A \textbf{112},
052211 (2025).

\bibitem{NAFTI01} T. Li and H. Hu, Floquet non-Abelian topological
insulator and multifold bulk-edge correspondence, Nat. Commun. \textbf{14},
6418 (2023).

\bibitem{NAFTI02} R.-J. Slager, A. Bouhon, and F. N. \"Unal, Non-Abelian
Floquet braiding and anomalous Dirac string phase in periodically
driven systems, Nat. Commun. \textbf{15}, 1144 (2024).

\bibitem{NAFTI03} J. Pan and L. Zhou, Floquet Non-Abelian Topological
Charges and Edge States, Chin. Phys. Lett. \textbf{42}, 090706 (2025).

\bibitem{GongPRX2018} Z. Gong, Y. Ashida, K. Kawabata, K. Takasan,
S. Higashikawa, and M. Ueda, Topological Phases of Non-Hermitian Systems,
Phys. Rev. X \textbf{8}, 031079 (2018).

\bibitem{AmoebaPRX2024} H.-Y. Wang, F. Song, and Z. Wang, Amoeba
Formulation of Non-Bloch Band Theory in Arbitrary Dimensions, Phys.
Rev. X \textbf{14}, 021011 (2024).

\bibitem{ETRMP} E. J. Bergholtz, J. C. Budich, and F. K. Kunst, Exceptional
topology of non-Hermitian systems, Rev. Mod. Phys. \textbf{93}, 015005
(2021).

\bibitem{MB1} Z.-H. Xu, X. Xia, and S. Chen, Exact mobility edges and topological phase transition in two-dimensional non-Hermitian quasicrystals, Sci. China Phys. Mech. Astron. {\bf 65}, 227211 (2022).

\bibitem{MB2} W. Han and L. Zhou, Dimerization-induced mobility edges and multiple reentrant localization transitions in non-Hermitian quasicrystals, Phys. Rev. B {\bf 105}, 054204 (2022).

\bibitem{MB3} L. Zhou and W. Han, Driving-induced multiple ${\cal PT}$-symmetry breaking transitions and reentrant localization transitions in non-Hermitian Floquet quasicrystals, Phys. Rev. B {\bf 106}, 054307 (2022).

\bibitem{MB4} R. Hamazaki, K. Kawabata, and M. Ueda, Non-Hermitian Many-Body Localization, Phys. Rev. Lett. {\bf 123}, 090603 (2019).

\end{thebibliography}
\end{document}